\documentclass[reprint,showpacs,showkeys,amsmath,amssymb]{revtex4-1}

\usepackage{amssymb}

\begin{document}

\title{The Variable-$c$ Cosmology\\ as a Solution to Pioneer Anomaly}

\author{Hossein Shojaie}
\email{h-shojaie@sbu.ac.ir}
\address{Department of Physics, Shahid Beheshti University, G.C., Evin, Tehran 1983963113, Iran}

\begin{abstract}
    It is shown that the Pioneer anomaly is a natural consequence of variable speed of light cosmological models wherein the speed of light is assumed to be a power-law function of the scale factor (or cosmic time). In other words, the Pioneer anomaly can be regarded as a non-gravitational effect of the continuously decreasing speed of light which indicates itself as an anomalous light propagation time delay in local frames. This time delay is accordingly interpreted as an additional Doppler blue shift.
\end{abstract}



\maketitle

\section{Introduction}

An unexpected Doppler blue shift in the radio signals reaching the earth from the Pioneer 10 \& 11 spacecrafts, when their distances from the Sun were between $20AU$ and $70AU$, shows a constant deceleration
\begin{equation}
A_P=(8.74\pm1.33)\times10^{-10}\frac{m}{s^2}\label{ap}
\end{equation}
towards the earth (or the sun), as reported first by~\citet{And98} and confirmed by more accurate data analysis~\citep{Tur06,Tur09,Tur11}. Since 1998, many explanations have been considered. These explanations can broadly be classified into two classes. The first corresponds to conventional physics including both on-board systematic effects~\citep{Ber07}, such as Thermal recoil force~\citep{Mur99,Kat99,Sch03,Ber08,Ber10,Tot09,Rie11}, and external forces, such as solar wind~\citep{And98}, effect of the expanding universe~\citep{Lam09},  and gravitational effects of mass distributions, mainly from Kuiper belt~\citep{Ber06,deD06,Nie06,Ior07} and interplanetary or interstellar dust~\citep{Nie05}. The second class corresponds to new physics, for instance, modified inertia~\citep{McC07}, dark matter distribution~\citep{Nie08}, cosmological constant~\citep{Mas08}, and modified (Hamiltonian) gravity~\citep{Ost02,Mof04,Ber04,Tre05,Jae05,Jae06,Bro06,Saf08,Avr09}.
But until now, none of them can explain this problem clearly. Moreover, as no firmly accepted anomaly has been detected in the orbits of planets yet~\citep{And02,Tan07,Ior10b}, it estranges the phenomenon to have a gravitational ground. In a review paper by~\citet{Nie07}, it has been claimed that the total estimated systematic error is about 10 percent of the observed un-modeled acceleration. So, it seems that the origin of this phenomenon may be a not-yet-discovered physical evidence. In addition, it should be mentioned that although some authors believe that the Pioneer anomaly cannot have a cosmological origin~\cite{Lam08,Miz05,Lac07}, some others argue that it \emph{can} be derived from conformal gravity considerations~\cite{Var10b,Var11}. It is worth noting that there is an approximate coincidence between the value of this anomaly and
\begin{equation}
H_0c_0\approx6.99\times10^{-10}\frac{m}{s^2}\label{H0C0},
\end{equation}
that is, the product of the Hubble constant and the speed of light, where the Hubble constant is assumed to be $H_0=100hKm/s/Mpc$ with $h\approx0.72$. For a detailed review of this phenomenon, one can see~\citet{Tur10} and references therein.

The aim of this work is to show that the Pioneer anomaly can be a natural consequence of those varying speed of light theories in which the speed of light is assumed to be a power-law function of cosmic time. The manuscript is arranged as follows. In Section~2, some arguments about the VSL theories are discussed. In Section~3, a VSL model provided by~\citet{Sho06,Sho07} is reviewed. According to this model, the Pioneer anomaly is a local effect of the universal variation of speed of light. This is done in Section~4. In Section~5, without considering a specific model, it is showed that one can use this anomaly to estimate the power of $c$, when it is supposed to be a power-law function of scale factor. Section~6 represents summary and remarks.

It is worth noting that beside the Pioneer anomaly, there are other more or less established anomalies within the solar systems such as the anomalous secular increase of the eccentricity of the lunar orbit~\citep{Ior11}, the anomalous precession of the perihelion of Saturn~\citep{Ior09} and the flyby anomaly~\citep{And07}. For a review of these anomalies see references~\citep{Lam08,And10} and references therein. Although these anomalies have not been asserted yet, they can be verified in the context of the VSL theory in further works.

\section{VSL Theories}

Varying speed of light theories, have been provided as alternatives to inflationary scenarios in order to solve the standard big bang (SBB) problems~\citep{Mof93,Bar98,Cla98,Bar99,AlbMag99,BarMag99,Alb99,
Cla00,Bas00,Mag00,You01,Mag03,Ell05, Ell07,MgM08,Ior10a}. As extensively discussed in the literature, the lack of a consistent field theory for VSL is a major difficulty of it. Besides, at least, two objections are against VSL theories.

The first is about the dimensionality of the speed of light, $c$. It is claimed that variations of dimensional quantities do not affect the physics and such variations can be removed by redefinitions of the units. However, any variation of a dimensionless quantity is a unit-free scalar, and since it is merely built up of dimensional parameters, its variation can be regarded as an indicator of physical (observer-independent) variation of one or more of its dimensional constituents. For instance, any probable variation of fine structure constant, $\alpha$, with respect to the cosmic time, is the indication of the actual time evolution of one or more fundamental constants that construct it, namely $c$, $\hbar$ or $e$. Moreover, there are many theories which regard variations of dimensional constants other than the speed of light, for example, scalar-tensor theories of gravitation~\citep{Fuj04} which concern probable variation of gravitational constant~\citep{Bra61}, or Beckenstein's proposal in which the fundamental electric charge has time-dependency~\citep{Bec82}.

The second objection is that the constancy of speed of light is one of the principles of the special relativity (SR). It can be shown that on the background of flat (Euclidean) geometry, relativity principle guaranties the existence of a frame-independent upper limit speed. If this speed is assumed to be infinite, then Galilean transformations and Newtonian absolute time arise. Otherwise, this invariant speed leads to Lorentz transformations and SR. In other words, the existence of an invariant upper limit speed in an Euclidean space-time, is a result of the relativity principle. Firm evidences, specially from electrodynamics, claim that this speed must be the speed of massless particles, or particularly the speed of light.

However, in the context of the whole universe, SR only defines local inertial frames. In other words, it is a local approximation of the global manifold obeying a gravitational theory such as general relativity (GR). This means that, although there is an invariant upper limit speed in each local inertial patch of the universe, which is ruled by SR, its value may differ from patch to patch, and any global gravitational theory which governs the whole universe may contain the derivatives of this locally invariant speed. So, a variable speed of light should enter in a gravitational theory minimally, in such a way that one can retrieve SR and Lorentz transformations in any local inertial frame. As an example, this can be easily done in a metric like
\begin{equation}
ds^2=c^2dt^2-g_{ij}dx^idx^j\label{102}\ ,
\end{equation}
where $g_{0i}=0$, by assuming $g_{00}=c^2(t)$ which consequently, adds a new non-zero connection component
\begin{equation}
\Gamma^0_{00}=\frac{\dot c}{c}\label{gamma000}\ .
\end{equation}
This component, like other connection components, is set to zero in any inertial frame, and one works with a constant $c$ in local frames.

Although the above example may be regarded as a minimal extension of SR to a generalized geometrical theory equipped with a variable $c$, simply replacing $c_0$, the coefficient of $dt$ in a metric by $c(t)$, is not enough to introduce a VSL model. Diffeomorphism tells us that this change does not contribute to a new physics and can be removed by redefinition of time coordinate. From Lagrangian viewpoint, terms with $\dot c$ which arise from metric, are absorbed to an exact differential term and will not cause any variation. This conclusion is valid until there is no other origin for $\dot c$ except metric. For instance, in a scalar-tensor theory of gravitation, provided the scalar field $\phi$ carries the dynamics of $c$ instead of $G$ (e.g. $\phi\equiv c^4$), $\dot c$ appears explicitly in the equations. (However, it should be mentioned here that a varying constant model is not automatically equivalent to a scalar-tensor theory.) It is worth noting that in a scalar-tensor theory, $\phi^{-1}$ is the coefficient which relates energy-momentum tensor $T_{\mu\nu}$ to the Einstein tensor $G_{\mu\nu}$, and determines the strength by which the ordinary matter affects geometry.

Switching back to the dynamics equations, namely the Einstein equation
\begin{equation}
G_{\mu\nu}=(8\pi G/c^4)T_{\mu\nu}\label{Einstein}
\end{equation}
when restricting ourselves to GR, the speed of light in the denominator of left-hand side, has not come from the metric and its time evolution has a physical significance. In addition, it is evident that in this way, $\dot c$ does not contribute to any vacuum solution of GR. This, for example, guarantees the static nature of the Schwarzschild solution.

Summarizing the above arguments, to construct a VSL model via generalizing GR, one should carry out all derivatives with respect to $dx^0$ (or $\partial x^0$) assuming $g_{00}=1$, and then replace $dx^0$ (or $\partial x^0$) with $cdt$ (or $c\partial t$) in the final relations wherever they appear provided the left-hand side of the Einstein equation is non-zero. This procedure can preserve the general covariance of a typical VSL theory and lets it reduce to SR in inertial frames. However, this procedure gives the same results as if one assumes $g_{00}=c^2(t)$ and $\Gamma^0_{00}=\dot c/c$ at first step.

\section{A variable-$c$ cosmological model}

Cosmological principle and Weyl postulate are of vital importance for the standard cosmology. These principles, which are in agreement with observations, are completely independent of GR. More precisely, the cosmological principle, namely the homogeneity and isotropy of the large scale universe, makes the Friedmann-Robertson-Walker (FRW) metric a preferable frame with least number of parameters which leads one to derive the Friedmann equations. On the other hand, by the Weyl postulate, the matter content of the universe can be chosen to be a perfect fluid.

In a VSL cosmological model, introduced by~\citet{Sho06,Sho07}, the universe is assumed to obey two additional principles, besides those accepted in GR. These two principles are:
\begin{description}
    \item[Principle(1):] The norm of a four-vector $(E,\textbf{p}c)$, is an invariant in any proper volume under general transformations. Specially, in FRW frame of cosmology, and according to the Weyl postulate and the property of the comoving frame, this norm reduces to
    \begin{equation}
    Mc^2={\rm const.}\label{ax1},
    \end{equation}
    in the proper volume within any arbitrary proper radius, where M is the enclosed rest mass in this volume.
    \item[Principle(2):] The relation
    \begin{equation}
    \frac{M}{Rc^2}={\rm const.}\label{ax2},
    \end{equation}
    holds in any proper volume with proper radius $R$. In other words, the ratio $M/Rc^2$ remains constant in any arbitrary proper volume.
\end{description}

According to this model~\citep{Sho07}, the universe begins from a hot big bang and expands and decelerates forever, like a flat model, despite the kind of its topology. That is, the flat universe, in contrast to Einstein-de Sitter models, is a stable case and, the Hubble parameter, $H(t)\equiv \dot a/a$ tends to zero az $t\rightarrow\infty$. The extra term related to the variation of the speed of light, $\dot c$, cancels out the effect of the curvature, $k$, in the dynamics equations. In addition, the principles lead matter content of the universe, if assumed to be a perfect fluid, to have the equation of state $p=-\frac{1}{3}\rho c^2$, similar to the value predicted from string models. It is worth mentioning that the speed of light in FRW frame varies as
\begin{equation}
c=c_0(\frac{a}{a_0})^{-\frac{1}{4}}=c_0(\frac{t}{t_0})^{-\frac{1}{5}}\label{VSL},
\end{equation}
where $t$ is the cosmic time. The detailed derivations and explanations can be found in Ref.~\citep{Sho07,Sho06}.

Other advantages of this model are that the flatness, the horizon and the relic abundances problems fade away from the SBB, without need to an add-on inflationary scenario. Also, it provides a better description about the fainter than expected flux of the supernovae type Ia via decreasing-$c$, instead of an accelerating epoch. In this case, one does not confront with more complicated subjects such as the cosmological constant problem. Moreover, the speed of light, $c$, can be regarded as a scalar field which decays continuously to produce matter in the universe.

\section{Pioneer Anomaly in VSL theory}

A key question should be answered here, before proceeding. Which metric determines the geodetic motion of a freely-falling object like the Pioneer spacecrafts, Schwarzschild metric or FRW metric? Nonetheless, the effect of the universal expansion is much smaller than the gravitational effects within galactic scales and this consequently shows that the geodesics are derived from a Schwarzschild metric (with the sun at the center). However, the effect of varying-$c$, at least far enough from the sun, may be significant merely by imposing an anomalous time delay. This can be seen, for example, by noting that the connection component $\Gamma^0_{00}$ can be no more zero when one deals with a VSL theory. A More detailed discussion is postponed to the Appendix.

To analyze the Pioneer anomaly in the VSL model described in the last section, let us consider the behavior of a signal from a spacecraft at comoving distance $r$ reaching the earth in FRW metric. Since the spacial distance between the spacecrafts and the sun are much more smaller than the size of the visible universe, one can neglect the effect of the curvature and simply assume $dr\approx dr/(1-kr^2)$.
Doing that, one has
\begin{equation}
r=\int_{t_e}^{t_0}\frac{c(t)}{a(t)}dt\label{800}
\end{equation}
where $t_e$ and $t_0$ denote the time of emission from the spacecraft and reaching the earth, respectively.
By expanding $c$ and $a$ up to first term around $t_0$ and using~(\ref{VSL}), one can write
\begin{eqnarray}
r&\approx&\int_{t_e}^{t_0}\frac{c_0+\dot c_0(t_0-t)}{a_0+\dot a_0(t_0-t)}dt\nonumber\\
&\approx&\int_{t_e}^{t_0}\frac{1}{a_0}\left[c_0-\frac{5}{4}H_0c_0(t_0-t)\right]dt\nonumber\\
&=& c_0\Delta t-\frac{5}{8}(H_0c_0)(\Delta t)^2\label{801}\ ,
\end{eqnarray}
where in the last step, it is assumed that $a_0=1$ and $\Delta t=t_0-t_e$. Obviously, the last term shows an extra unexpected term that can be explained as a negative constant acceleration, $A$, which value is
\begin{equation}
A=\frac{5}{4}H_0c_0\approx 8.73\times10^{-10}\frac{m}{s^2}\label{802}
\end{equation}
in a very good agreement with the Pioneer anomaly, where~(\ref{H0C0}) is used.
In deriving this result, the way the object moves does not affect the result. It is completely independent of the velocity, acceleration and direction of the motion, and instead, depends on the Hubble expansion rate and the speed of light. The direction of the deceleration is always towards the observer, that is, in direction of the signal ray. Moreover, as this deceleration is related to the receiving signal and not to the object itself, it does not contribute to any precession in the orbits of the planets.

\section{Conversed View}

There can be a conversed approach to this problem in VSL context, but this time, without assuming any specific model. Supposing the validity of the cosmological principle, one simply ends with FRW metric~(\ref{102}). Moreover, if one assumes that the speed of light in this cosmological frame is a power-law function of the scale factor, that is $c\propto a^n$, then one can estimate the range of $n$ by derivation method similar to~(\ref{801}), using~(\ref{ap}) and~(\ref{H0C0}). This means that the relation~(\ref{801}) is modified to
\begin{eqnarray}
r&\approx&\int_{t_e}^{t_0}\frac{1}{a_0}\left[c_0+(n-1)H_0c_0(t_0-t)\right]dt\nonumber\\
&=& c_0\Delta t+\frac{n-1}{2}H_0c_0(\Delta t)^2\label{803}
\end{eqnarray}
up to first order. This evidently implies a constant deceleration
\begin{equation}
A=-(n-1)H_0c_0\label{804}\ .
\end{equation}
Equating~(\ref{804}) and~(\ref{ap}), and using~(\ref{H0C0}) then leads to
\begin{equation}
n=-0.25\pm0.19\label{n}\ .
\end{equation}
Not surprisingly, equation~(\ref{n}) is a manifestation of decreasing-$c$ in an expanding universe.

\section{Summary and Remarks}

The Pioneer anomaly is an un-modeled Doppler blue shift, equivalent to a constant deceleration, which has not had a conventional physical explanation yet. In this work, it has been shown that a VSL cosmological model can generally predict this anomaly, provided that firstly the cosmological principle applies and secondly the speed of light is a power-law function of the scale factor, with its power lying in the range $(-0.44,-0.06)$.

According to this scenario, it has been argued that this negative acceleration is not due to a true acceleration, but merely is an anomalous light propagation time delay, interpreted as an unexpected blue shift. Hence, it does not depend on the velocity, acceleration or direction of the motion of the object. The negativity of this anomaly implies that the Hubble constant and $\dot c$ have opposite signs. That is, in an expanding universe, the speed of light is decreasing.

A complaint against the procedure followed in~(\ref{801}) and~(\ref{803}), may be the issue of embedding a central body solution, that is a Schwarzschild metric, within an overall cosmological background, namely the FRW metric. The situation somehow is similar to considering local inertial frames of freely-falling objects within the background of a Schwarzschild metric; the freely-falling observer cannot detect gravity, but the tidal force can be observed by him/her. The Pioneer anomaly can be similarly regarded as a local detectable effect of a universal evolution of the speed of light. Nonetheless, the author believes that this phenomenon is the only evidence that exhibits the expansion of the universe locally.

\section*{Acknowledgment}

The author would like to appreciate Dr. Mehrdad Farhoudi for his hospitality and useful comments. This work has been financially supported by the research council of Shahid Beheshti University.

\section*{Appendix}

There are three main differences between the Schwarzschild metric and FRW metric, which make the combination of these two metrics difficult:
\begin{itemize}
  \item The Schwarzschild metric is static but FRW is a dynamical metric.
  \item The Schwarzschild metric is a vacuum solution while FRW, at least when applied to the real universe, concerns the perfect fluid as its energy-momentum tensor.
  \item The Schwarzschild metric is inhomogeneous since it describes the spacetime outside a spherically symmetric object. On the other hand, homogeneity is one of two assumptions in deriving FRW metric.
\end{itemize}

However,to describe the evolution of black holes and other local systems such as the solar system, collapsing stars, galaxies, clusters and super clusters of galaxies in an expanding universe, one needs a metric which asymptotically tends to FRW metric at large radius while at the same time reduces to Schwarzschild spacetime at significantly small distances. In addition, the conjunction conditions between two metrics, according to each specific problem, should be appropriately applied. Attempts to obtain such a solution has been dated back to 1933 by~\citet{McV33} and followed by the others~\citep{Ein45,Pac63,Nol98}. But until now, there is no definitive and well-defined metric for this purpose and all attempts have failed.

Despite the above fact, the McVittie's metric seems to be the best prototype for embedding a spherically symmetric object in an expanding universe. In its general form, it is written as
\begin{equation}
ds^2=\left(\frac{1-\mu(t,r)}{1+\mu(t,r)}\right)^2c^2dt^2-\left(1+\mu(t,r)\right)^4a^2(t)\frac{dr^2+r^2d\Omega^2}{(1+\frac{1}{4}kr^2)^2}\label{A1}\ ,
\end{equation}
where
\begin{equation}
\mu(t,r)=\frac{Gm}{2a(t)rc_0^2}\sqrt{1+\frac{1}{4}kr^2}\label{A2}\ .
\end{equation}
The constant parameters $m$ and $k$ can be related to the mass of the central object and the curvature of the space, respectively, and $a(t)$ is the asymptotic cosmological scale factor. The solution is exactly FRW for $m=0$ and wholly Schwarzschild metric when $k=0$ and $a=1$. Moreover, for $\mu\ll1$ the metric reduces to
\begin{equation}
ds^2=(1+2\mu)c^2dt^2-(1-2\mu)a^2(t)\left(dr^2+r^2d\Omega^2\right)\label{A3}\ ,
\end{equation}
which is obviously the perturbed FRW metric in Newtonian gauge, that is, $\mu$ can be regarded as the Newtonian gravitational potential. For a sun-size star, this perturbed metric is well worth applicable outside the star; at much larger distances, it is fully FRW metric. It can be easily shown that according to metric~(\ref{A1}) (and consequently the perturbed metric~(\ref{A2})), six components of connections (excluding those which are the same in both metrics), namely,
\begin{equation}
\Gamma^1_{01}=\Gamma^1_{10}=\Gamma^2_{02}=\Gamma^1_{20}=\Gamma^3_{03}=\Gamma^1_{30}=\frac{\dot a}{a}\label{A4}\ ,
\end{equation}
are not negligible even for small $\mu$. If one also assumes that the speed of light is a function of $t$, then in addition to the above connection components,
\begin{equation}
\Gamma^0_{00}=\frac{\dot c}{c}\label{A5}
\end{equation}
cannot be disregarded for small values of $\mu$, too. That is, concerning the McVittie's metric, these components of connection are capable of bringing the effect of global expansion to local systems. It should be mentioned that, there have always been doubts about the lowest scale at which expansion can be observed. As it has been stated in the literature~\citep{And95}, in principle, there is no restriction about this scale and the expansion proceeds in all scales. However, in practice, local systems do describe the geometry and consequently there is a scale that beyond it objects do not feel expansion.

The lack of an appropriate metric in this situation, leads authors to consider the effect of universal expansion in local systems in a different way. In this method, the local system is considered in a local inertial frame and the effect of the background FRW metric is shown in deviation from the geodetic motion. In this sense, and using the relation
\begin{equation}
\frac{d^2x^i}{dt^2}+\Gamma^i_{jk}\frac{dx^j}{dt}\frac{dx^k}{dt}=-{R^i}_{0j0}x^j\label{A6}\ ,
\end{equation}
one has
\begin{equation}
\frac{d^2r}{dt^2}-\left(\frac{\ddot a}{a}-\frac{\dot c}{c}\frac{\dot a}{a}\right)r=0\label{A7}\ .
\end{equation}
Obviously, the third term in the above relation is due to VSL theory, that is, an additional deviation from the standard geodesic deviation (second term). Not surprisingly, $\ddot a$ is also affected by varying $c$. Adding the gravitational effect of the sun, relation~(\ref{A7}) becomes
\begin{equation}
\frac{d^2r}{dt^2}-\left(\frac{\ddot a}{a}-\frac{\dot c}{c}\frac{\dot a}{a}\right)r+\frac{Gm}{r^2}-\frac{L_o^2}{m_or^3}=0\label{A8}\ ,
\end{equation}
where $m_o$ and $L_o$ are the mass and the angular momentum of the object, respectively. Actually, this deviation in a local inertial frame is a light propagation time delay in FRW, which is shown in the text.

It should be mentioned that a natural connection can be seen between Schwarzschild and FRW metrics in the context of conformal gravity~\cite{Man89,Man06,Var10a} where appropriate conformal transformations can smoothly interpolate between the two metrics.

\bibliographystyle{elsarticle-harv}

\end{document}